\documentclass[conference]{IEEEtran}

\usepackage{etoolbox}

\makeatletter  
  {\scshape}
  {}
  {}
  {}
\makeatletter
\patchcmd{\@makecaption}
  {\\}
  {.\ }
  {}
  {}
\makeatother

\usepackage{amsfonts}
\usepackage{mathrsfs}
\usepackage{amsfonts}
\usepackage{amssymb}
\usepackage{graphicx}
\usepackage{epsfig}
\usepackage{epstopdf}
\usepackage{psfrag}
\usepackage{amsmath}
\usepackage{array}
\usepackage{cases}
\usepackage{longtable}
\usepackage{subfigure}
\usepackage{cite,graphicx,amsmath,amssymb,color}
\usepackage{anysize}
\usepackage{algorithmic}
\usepackage{algorithm}
\usepackage{caption}

\newcommand{\dis}{\stackrel{d}{\sim}}

\newtheorem{Thm}{Theorem}
\newtheorem{Lem}{Lemma}

\marginsize{13mm}{13mm}{19mm}{43mm}
\IEEEoverridecommandlockouts

\begin{document}
\title {Temporal-Spatial Aggregation for \\ Cache-Enabled Wireless Multicasting Networks \\ with Asynchronous Content Requests}

\author{\IEEEauthorblockN{Jifang Xing, \ Ying Cui}\IEEEauthorblockA{Shanghai Jiao Tong University, China}\and \IEEEauthorblockN{Vincent Lau}\IEEEauthorblockA{HKUST, Hong Kong}}

\maketitle

\begin{abstract}
Existing multicasting schemes for massive content delivery do not fully utilize multicasting opportunities in  delay tolerant content-oriented applications. In this paper, we propose a novel temporal-spatial aggregation-based  multicasting scheme  in a large-scale cache-enabled wireless network.
The proposed scheme can efficiently exploit  multicasting opportunities in asynchronous content requests to improve spectral efficiency.
By making use of the delay tolerance of elastic services, the proposed scheme achieves a better energy-throughput-delay tradeoff.
Utilizing tools from stochastic geometry, we derive a tractable expression for the successful transmission probability in the general region.
Using asymptotic approximations, we derive closed form successful transmission probabilities in the large delay region as well as the large and small user density regions.
The asymptotic results reveal that the successful transmission probability  increases and the energy consumption decreases at the cost of delay increase in these asymptotic regions. The analysis in this paper provides a new understanding of the energy-throughput-delay tradeoff for massive content delivery in  large-scale cache-enabled wireless networks.
\end{abstract}

\section{Introduction}\label{sec:intro}

Mobile data traffic has been shifting from connection-oriented services such as voice telephony and text messaging to content-oriented services such as media streaming, application downloads/updates and social network communications.
Regarding content-oriented services, there has been an increasing demand to send the same popular contents to multiple users.
Owing to the broadcast nature of the wireless medium, multicasting is an efficient way to deliver widely popular contents from a transmitter to multiple  users concurrently.
A lot of existing works on multicasting have assumed that all the content requests arrive synchronously to the same cell  and focus on the analysis of multicasting gains under this assumption \cite{kim2008ofdma,wang2010throughput}.
To fully unleash  the multicasting benefit, one needs to accommodate asynchronous and spatially distributed requests in wireless networks.

Caching popular contents at base stations (BSs) is also an effective way for massive content delivery in wireless networks, as it can greatly alleviate backhaul load and reduce service time. Multicasting and caching have been jointly considered for content-oriented applications in wireless networks\cite{huang2016delay,poularakis2016exploiting,cui2015analysis}.
For instance, in \cite{huang2016delay,poularakis2016exploiting}, the authors consider multicasting to minimize the average energy and delay costs for elastic services in cache-enabled wireless networks, by aggregating asynchronous  common requests within a fixed time window.
The temporal aggregations of asynchronous common content requests yield a tradeoff between the spectral efficiency gain due to increased multicasting opportunities  and the access delay penalty due to temporal aggregations.
Note that the
pure temporal aggregations of common content requests at the same cell in \cite{huang2016delay,poularakis2016exploiting} may not fully exploit multicasting opportunities because there are a lot of  common requests which are generated at different spatial locations.
In  \cite{cui2015analysis}, the authors consider spatial aggregations of synchronous common content requests at nearby cells, without considering temporal  aggregations of asynchronous common content requests.
In addition,  in the analysis and design, \cite{huang2016delay,poularakis2016exploiting} fail to consider the stochastic natures of channel fading and  geographic locations of BSs and users  as well as interference, which are  key aspects in wireless networks.

In this paper, we consider content-centric applications with asynchronous and spatially distributed content requests for  elastic services in  a large-scale cache-enabled wireless network.
We propose a novel temporal-spatial aggregation-based multicasting scheme to enhance multicasting opportunities  so as to improve spectral efficiency.
The proposed multicasting scheme creates more multicasting opportunities at the same energy and delay costs compared with the existing  multicasting schemes in \cite{huang2016delay,poularakis2016exploiting}, and hence it can achieve a better energy-throughput-delay tradeoff.
By carefully handling different types of interferers and adopting appropriate approximations, we derive a tractable expression for the successful transmission probability in the general region, utilizing tools from stochastic geometry.
Using asymptotic approximations, we derive closed form successful transmission probabilities in  the large delay region as well as the large and small user density regions, respectively.
The asymptotic results reveal that the successful transmission probability  increases and the energy consumption decreases at the cost of delay increase  in these asymptotic regions.
The analysis in this paper offers a new understanding of the energy-throughput-delay tradeoff for massive content delivery in  large-scale cache-enabled wireless networks.

\section{System Model}

\subsection{Network Model}

\begin{figure}
 \begin{center}
 \subfigure[Legends]
 {\resizebox{2.8cm}{!}{\includegraphics{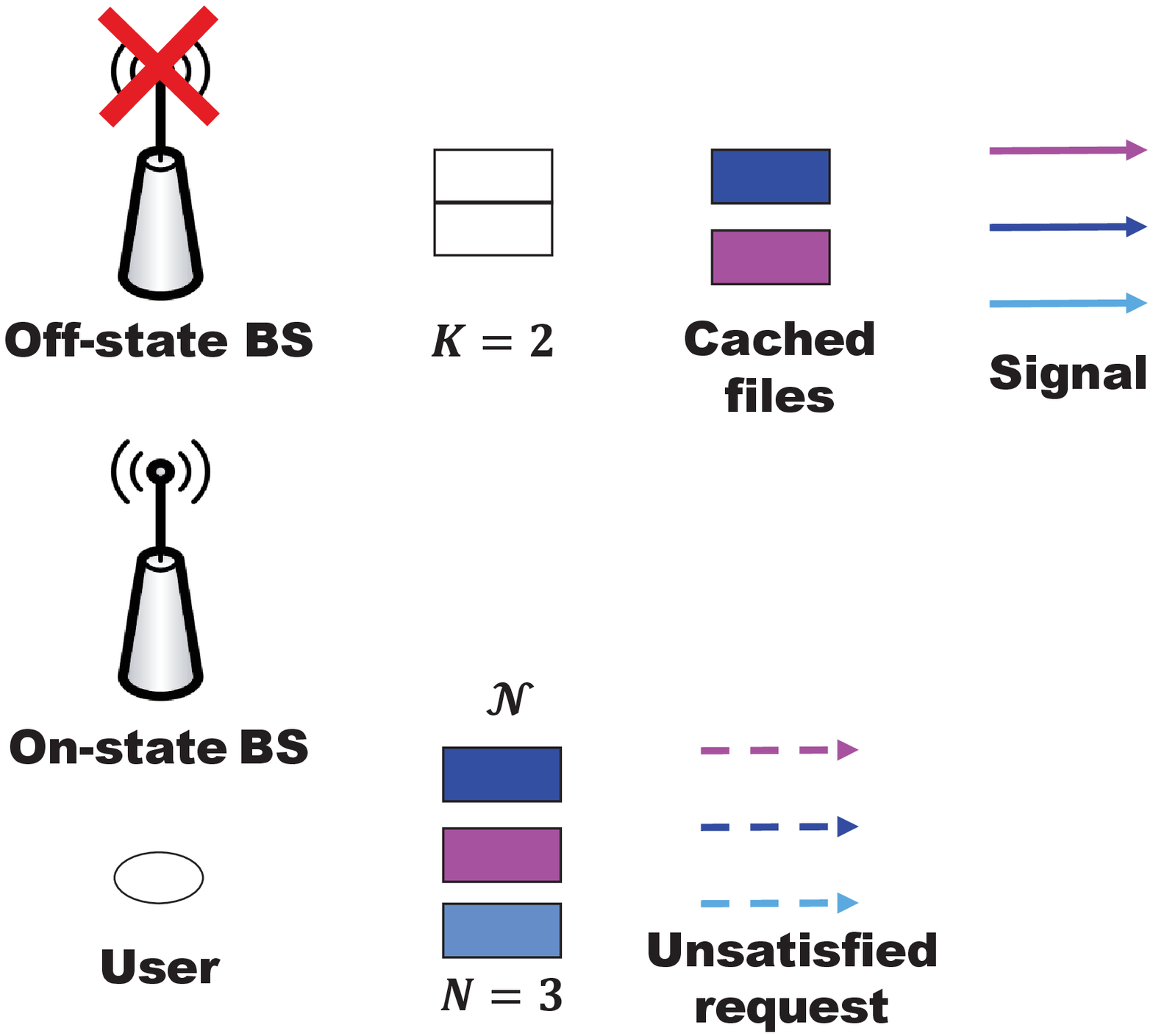}}}
 \subfigure[$t=1$]
 {\resizebox{2.8cm}{!}{\includegraphics{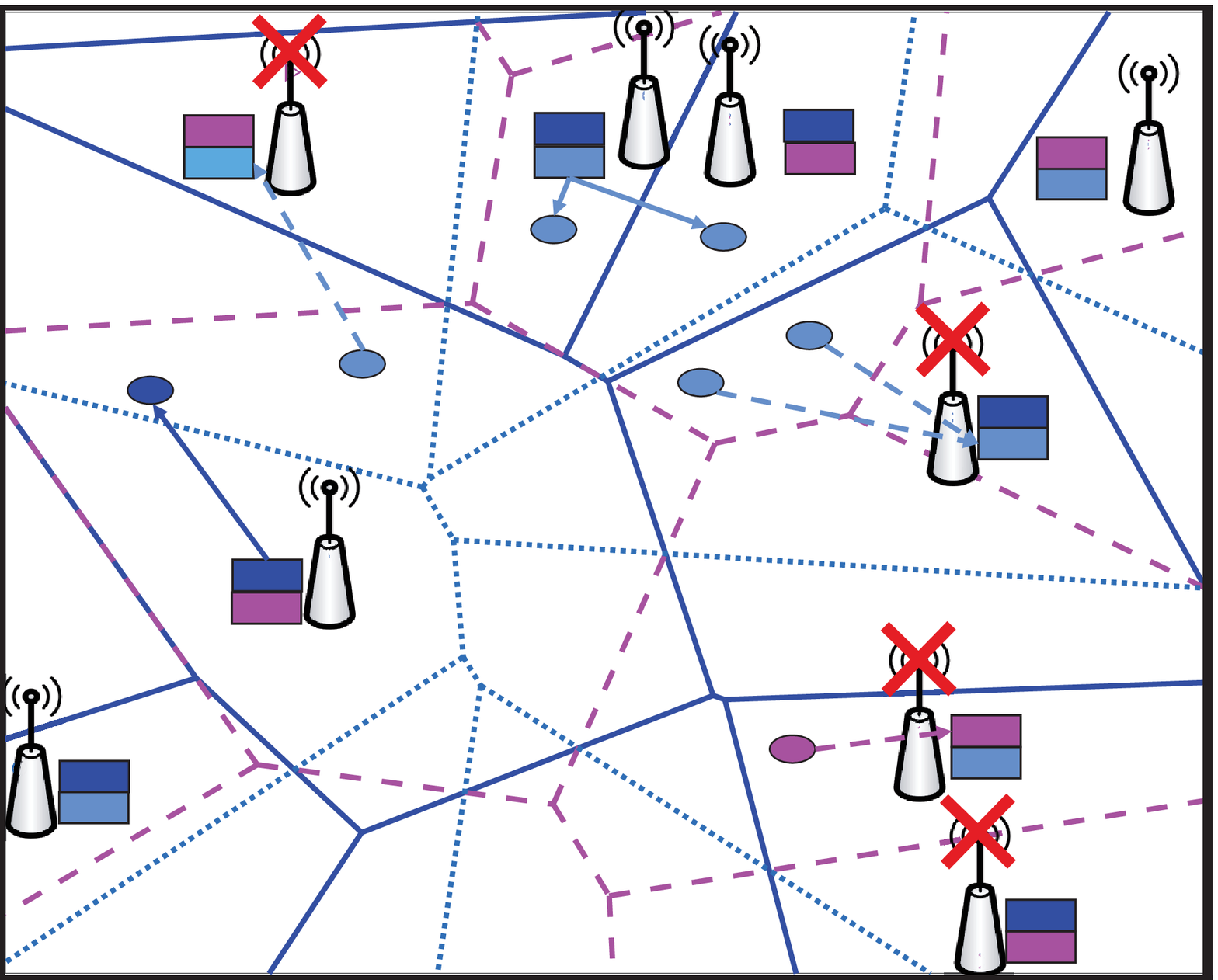}}}\\
  \subfigure[$t=2$]
 {\resizebox{2.8cm}{!}{\includegraphics{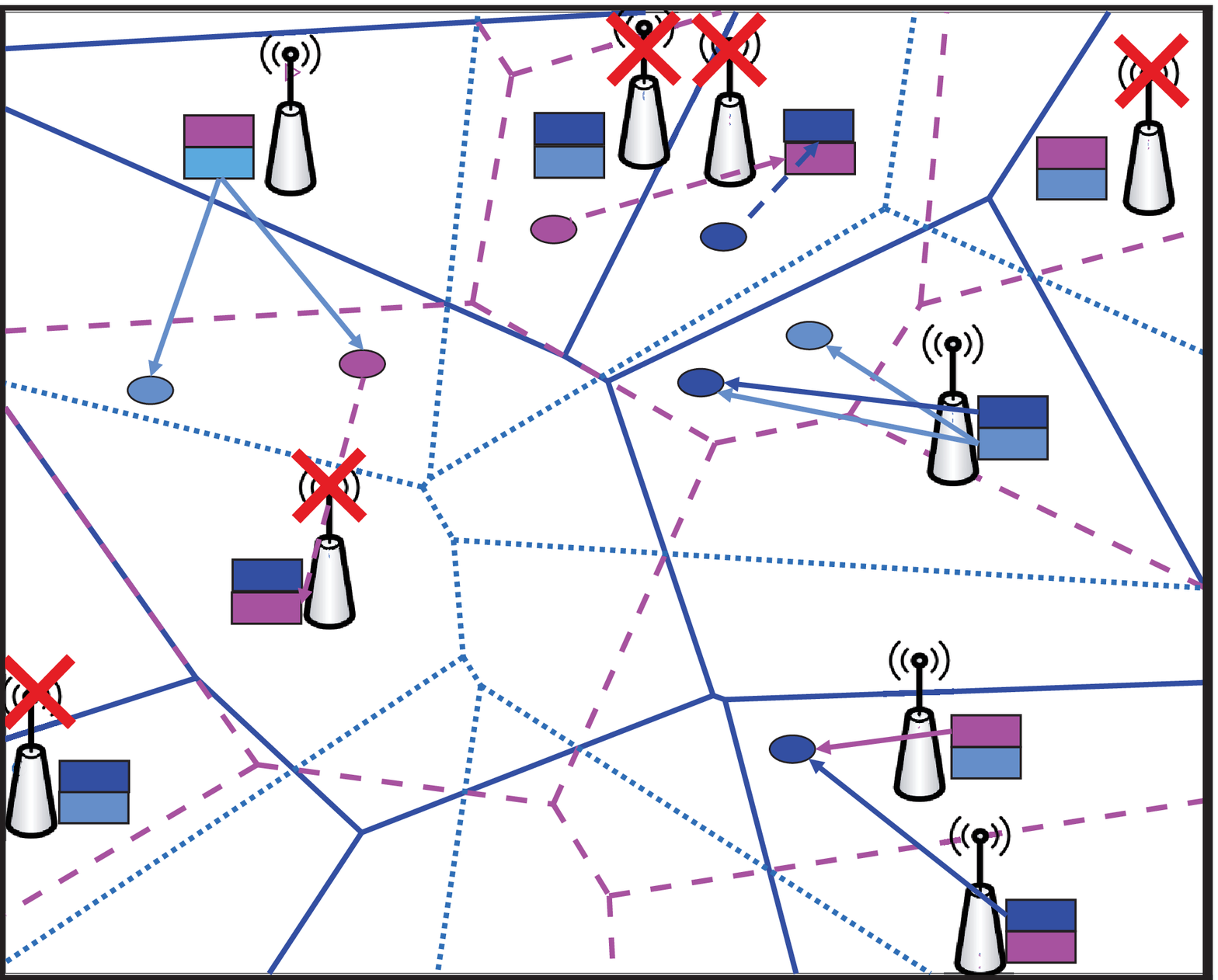}}}
  \subfigure[$t=3$]
 {\resizebox{2.8cm}{!}{\includegraphics{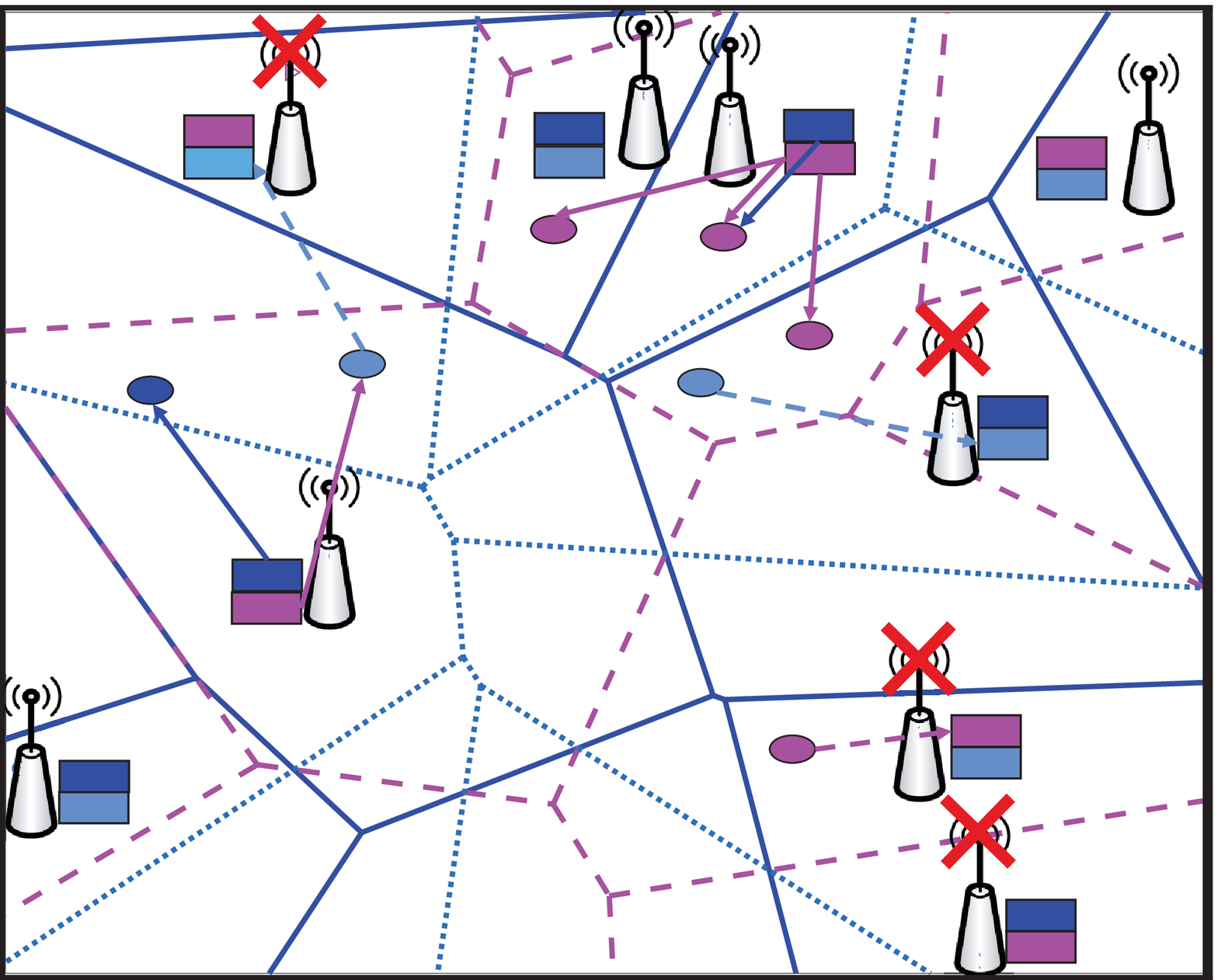}}}
 \end{center}
 \caption{Illustration example ($T=2$).}\label{fig:system}
\end{figure}

As illustrated in Fig~\ref{fig:system}, we consider a large-scale network model.\footnote{The network model and random caching design in this paper are similar to those in \cite{cui2015analysis}.
The multicasting design and performance analysis in this paper are more general and include those in \cite{cui2015analysis} as special cases.}
The locations of BSs are spatially distributed as an independent homogeneous Poisson point process (PPP) $\Phi_b$ with density $\lambda_b$.
The locations of users are distributed as an independent homogeneous PPP $\Phi_u$ with density $\lambda_u$.
We consider the downlink transmission.
Each BS has one transmit antenna with transmission power $P$.
Each user has one receive antenna.
The total bandwidth is $W$ (Hz).
Consider a discrete-time system with time being slotted.
Let $t= 1,2,\ldots$ denote the slot index.
Due to large-scale pathloss, transmitted signals with distance $D$ are attenuated by a factor $D^{-\alpha}$, where $\alpha >2$ is the path loss exponent.
For small-scale fading, we assume Rayleigh fading. 
Let ${\mathcal N}\triangleq \{1,2,\ldots,N\}$ denote the set of $N\geq 1$ files (contents) in the network.
For ease of illustration, we assume that all  files  have the same size.
Each file is of certain popularity.
We assume that  the file popularity distribution is identical among all users.
At the beginning of each slot, each user randomly  requests one file which is file $n\in \mathcal N$ with probability $a_n\in (0,1)$, where $\sum_{n\in \mathcal N}a_n=1$.
Thus, the file popularity distribution is given by $\mathbf a\triangleq (a_n)_{n\in \mathcal N }$, which   is assumed to be known priori.
In addition, without loss of generality (w.l.o.g.), we assume $a_{1}\ge a_{2}\ldots\ge a_{N}$. We consider delay tolerant services. Each file request may not need to be satisfied within one slot.

\subsection{Random Caching}

The network consists of  cache-enabled BSs.
In particular, each BS is equipped with a cache of size $K\leq N$, storing  $K$  different  files out of $N$.
Every $K$ different files form a combination.
There are $I\triangleq \binom{N}{K}$ different combinations of $K$  different  files in total.
Let $\mathcal I$ denote the set of $I$ combinations.
Combination $i\in \mathcal I$ can be characterized  by an $N$-dimensional vector $\mathbf x_i\triangleq (x_{i,n})_{n\in \mathcal N}$, where $x_{i,n}=1$ indicates that file $n$ is included in combination $i$ and $x_{i,n}=0$ otherwise.
Denote $\mathcal N_i\triangleq \{n:x_{i,n}=1\}$ as the set of $K$   files contained in combination $i$.
We consider random caching on the basis of file combinations.
Each BS cache stores combination $i\in \mathcal I$ with probability $p_i$ satisfying $0\leq p_i\leq1,\ i\in \mathcal I$ and $\sum_{i\in \mathcal I}p_i=1$.
A random caching design is specified by the caching distribution $\mathbf p\triangleq (p_i)_{ i\in \mathcal I}$.
Let $\mathcal I_n\triangleq \{i\in \mathcal I:x_{i,n}=1\}$ denote the set of $I_n\triangleq \binom{N-1}{K-1} $ combinations  containing file $n$.
Let $T_n\triangleq \sum_{i\in\mathcal{I}_{n}}p_{i}$ denote the probability that file $n$ is stored at a BS.
In this paper, we focus on serving  cached files at  BSs to get first-order insights into the design of cache-enabled wireless networks.
BSs may serve uncached files through other service mechanisms, the investigation of which is beyond the scope of this paper.

\subsection{Temporal-Spatial Aggregation-Based Multicasting}\label{Sec:multi and onoff}

As discussed in Section~\ref{sec:intro}, content requests are usually generated at different times and  locations.
To fully unleash  the multicasting benefit, we propose a novel temporal-spatial aggregation-based multicasting scheme where the  aggregation of requests in time is  based on a periodic BS on/off mechanism and the aggregation in space is induced by  caching. Specifically,
we  consider a periodic BS on/off mechanism with period $T\ge1$ (in slots). First, all BSs are randomly divided into $T$ tiers according to the uniform distribution.
That is, a BS is assigned to tier $\tau\in\{1,\ldots,T\}$ with probability $1/T$.
The locations of the BSs in each tier $\tau$ are distributed as an independent homogeneous PPP $\Phi_{b,\tau}$ with density $\lambda_u/T$.
Note that each tier of BSs is on once every $T$ slots.
In particular, if $t\mod T=\tau$, then only the BSs in tier $\tau$ are on in slot $t$.
An active BS (being on) can serve file requests, while an inactive BS (being off) cannot.
We assume a BS can always receive file requests, no matter being on or off.
In slot $t$, a user requesting file $n$ submits the file request to the neatest BS storing a combination $i\in \mathcal I_n$, referred to as the serving BS of this file request, as this BS offers the maximum long-term average receive power for file $n$ at this user.
This file request will be served by this BS within $T$ slots once the BS is on.
Note that the serving BS of a user's file request may not be its geographically nearest BS and is also statistically determined by the caching distribution $\mathbf p$.

We propose a temporal-spatial aggregation-based multicasting scheme for delay tolerant content dissemination in the cache-enabled wireless network.
Consider a BS which is on in slot $t\geq T$ (and is off in the previous $T-1$ slots).
Let $K_0\in \{1,2,\ldots,K\}$ denote the number of different file requests for the $K$ files stored at this BS during the latest $T$ slots, including the current slot and the previous $T-1$ slots.
In slot $t$, the active BS transmits each of the $K_0$ files at rate $\theta$ (bits/second) and over $\frac{1}{K_0}$ of the total bandwidth $W$ using frequency division multiple access (FDMA).
In other words, in slot $t$, the active BS tries to serve all the users requesting the same file from it in the latest $T$ slots using one single transmission, and these users try to decode the file from the single transmission.
Note that  the proposed multicasting scheme  with $T=1$ reduces to the continuous multicasting scheme in \cite{cui2015analysis}, where each BS is always on and serves file requests in every slot without considering temporal aggregations of asynchronous content requests.

The proposed temporal-spatial aggregation-based multicasting scheme with periodic BS on/off of period $T$ exhibits the following features, which cannot be realized in the existing multicasting schemes for cache-enabled wireless networks.
\begin{itemize}
  \item {\bf Reduction of Energy Consumption (and Interference):} The transmission energy consumption (and interference) is only caused by approximate $1/T$ of BSs, and hence the energy consumption (and interference) is greatly reduced,   \item {\bf Increase of Multicasting Opportunities:} Each active BS serves $T$ times of the file requests in one slot, and hence the chance of a transmitted file being requested by more than one user significantly increases.
\end{itemize}

Therefore, by temporal aggregation,  this multicasting scheme can achieve $1/T$ energy consumption and $T$ times multicasting opportunities, at the cost of $T$ times delay,  compared to the continuous  multicasting scheme in \cite{cui2015analysis}. In addition, by spatial aggregation, this multicasting scheme can achieve more multicasting opportunities at the same energy and delay costs as the discrete multicasting schemes in \cite{huang2016delay,poularakis2016exploiting}.
In the following, we mainly focus on  investigating  how the reduction of the energy consumption and the increase of the multicasting opportunities affect the content delivery.

\section{Performance Metric}

In this paper, w.l.o.g., we study the performance for serving a file request from a typical user $u_0$ located at the origin at a typical slot $t_0\geq T$.
For ease of analysis, we assume all BSs are active.
Supposing the file requested by $u_0$ at slot $t_0$ is file $n$,
file $n$ is transmitted by $B_{n,0}\in \Phi_{b,\tau_0}$ in tier $\tau_0$ at slot $t_{0}^{'}$ ($t_{0}^{'}=T\left\lceil\frac{t_0-\tau_0}{T}\right\rceil+\tau_0$).
Then, at slot $t_{0}^{'}$, the corresponding received signal at $u_0$ is given by $y_{0}=D_{0,0}^{-\frac{\alpha}{2}}h_{0,0}x_{0}+\sum_{\ell\in\Phi_{b,\tau_{0}}\backslash B_{n,0}}  D_{\ell,0}^{-\frac{\alpha}{2}}h_{\ell,0}x_{\ell}+n_{0},
$
where $D_{0,0}$ is the distance between $u_{0}$ and  $B_{n,0}$, $h_{0,0}\dis \mathcal{CN}\left(0,1\right)$ is the small-scale channel between $B_{n,0}$ and $u_{0}$, $x_{0}$ is the transmit signal from $B_{n,0}$  to $u_{0}$, $D_{\ell,0}$ is the distance between BS $\ell$ and $u_{0}$, $h_{\ell,0}\dis \mathcal{CN}\left(0,1\right)$ is the small-scale channel  between BS $\ell$ and $u_{0}$, $x_{\ell}$ is the transmit signal from BS $\ell$  to its scheduled users,  and $n_{0}\dis\mathcal{CN}\left(0,N_{0}\right)$ is the complex additive white Gaussian noise of power $N_0$.
The  signal-to-interference plus noise ratio (SINR) for $u_{0}$ at slot $t_0^{'}$ is given by ${\rm SINR}_{n,0} = \frac{{D_{0,0}^{-\alpha}}\left|h_{0,0}\right|^{2}}{\sum_{\ell\in\Phi_{b,\tau_{0}}\backslash B_{n,0}}D_{\ell,0}^{-\alpha}\left|h_{\ell,0}\right|^{2}+\frac{N_{0}}{P}}$. 
Note that when $K> 1$, ${\rm SINR}_{n,0}$ is a continuous random variable, the probability density function (p.d.f.) of which depends on the BS on/off period $T$.
In addition, at slot $t_{0}^{'}$, let $K_{n,0}\in \{1,\cdots, K\}$ denote the number of requests for different files received by the users from $B_{n,0}$ during the latest $T$ slots, indicating the file load  of BS $B_{n,0}$ at slot $t_{0}^{'}$.
Note that when $K> 1$, $K_{n,0}$ is a discrete random variable, the probability mass function (p.m.f.) of which depends on the user density $\lambda_u$ and BS on/off period $T$.
Under the proposed temporal-spatial aggregation-based multicasting scheme, each of the $K_{n,0}$ requested files is sent over bandwidth $\frac{W}{K_{n,0}}$ to all the users requesting the file from $B_{n,0}$ within the latest $T$ slots.
Thus, the capacity of the channel between $B_{n,0}$ and $u_0$ at slot $t_{0}^{'}$ is given by $C_{n,K,0}\triangleq \frac{W}{K_{n,0}}\log_{2}\left(1+{\rm SINR}_{n,0}\right)$.
The dissemination of file $n$ at rate $\theta$ can be decoded correctly at $u_0$ if $C_{n,K,0}\geq \theta$.
Then, the successful transmission probability of  file $n$ requested by $u_0$ at slot $t_0$ is given by\footnote{Please note that the distributions of random variables $K_{n,0}$ and ${\rm SINR}_{n,0}$ depend on $\mathbf p$ and $T$.
Thus, we write $ q_{n}(\mathbf p,T)$ as a function of $\mathbf p$ and $T$.
}
 \begin{align}
 q_{n}(\mathbf p,T)\triangleq{\rm Pr}\left[\frac{W}{K_{n,0}}\log_{2}\left(1+{\rm SINR}_{n,0}\right)\geq\theta\right].\label{eqn:succ-prob-n-def}
 \end{align}
Requesters are mostly concerned about whether  their desired files can be successfully received.
Therefore, in this paper, we consider the successful transmission probability  of a file request from $u_0$ at slot $t_0$, also referred to as successful transmission probability, as the network performance metric.
By total probability theorem,  the successful transmission probability   under the proposed scheme is given by $q({\mathbf p},T)\triangleq\sum_{n\in \mathcal N}a_{n}q_{n}(\mathbf p, T).$

\section{Performance Analysis In General Region}

\begin{figure*}[!t]
\begin{center}
    \subfigure[\label{fig:pro_tau}$q({\mathbf p},T)$ versus $\theta$. $T_1=2$, $T_2=1$, $\lambda_u=0.1$ , $\frac{P}{N_0}=30\;{\rm dB}$.]
  {\resizebox{5cm}{!}{\includegraphics{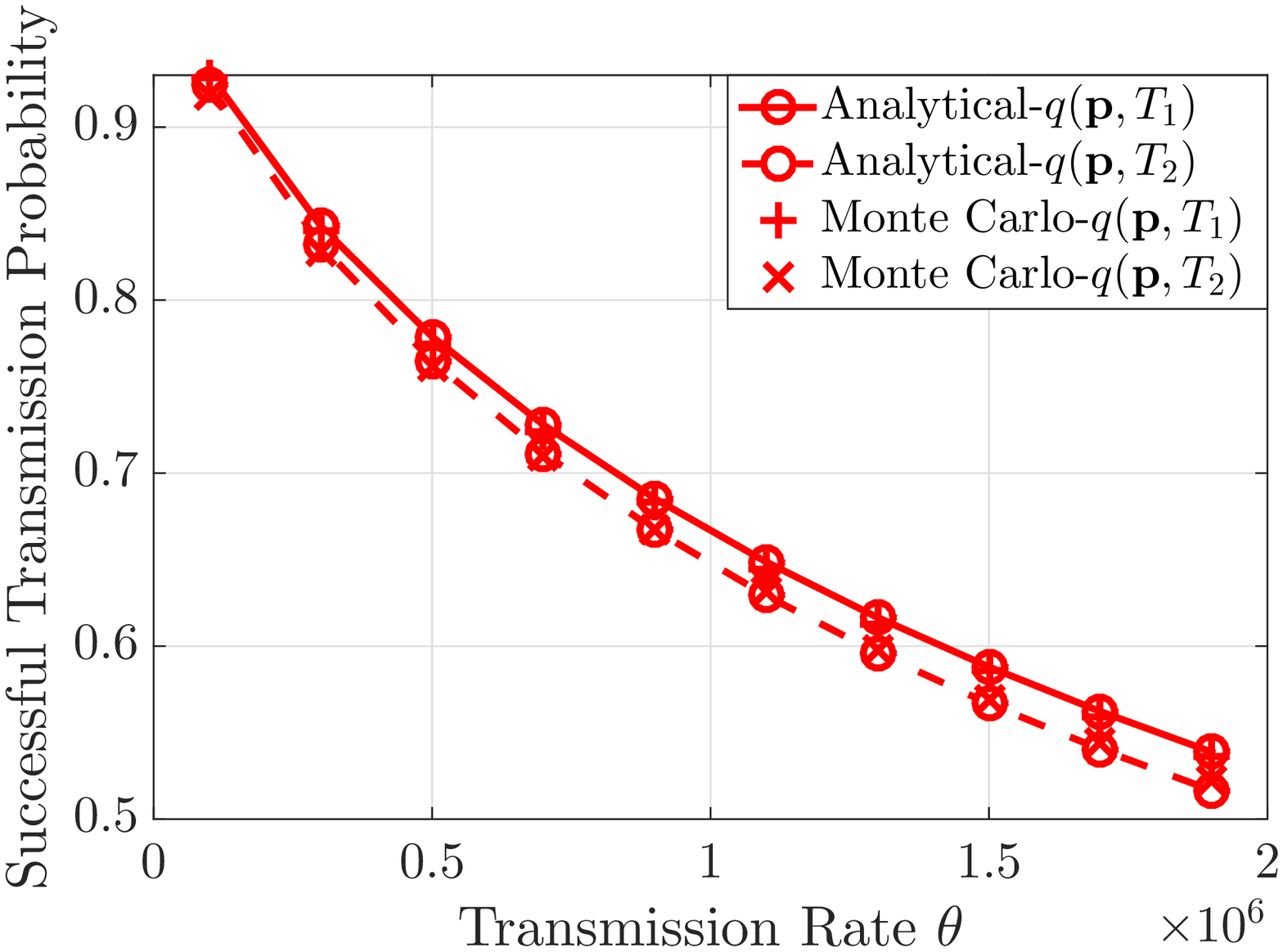}}}\quad\quad
            \subfigure[\label{fig:baselines} Performance comparison. $T=2$, $\lambda_u=0.1$, $\frac{P}{N_0}=30\;{\rm dB}$.]
  {\resizebox{5cm}{!}{\includegraphics{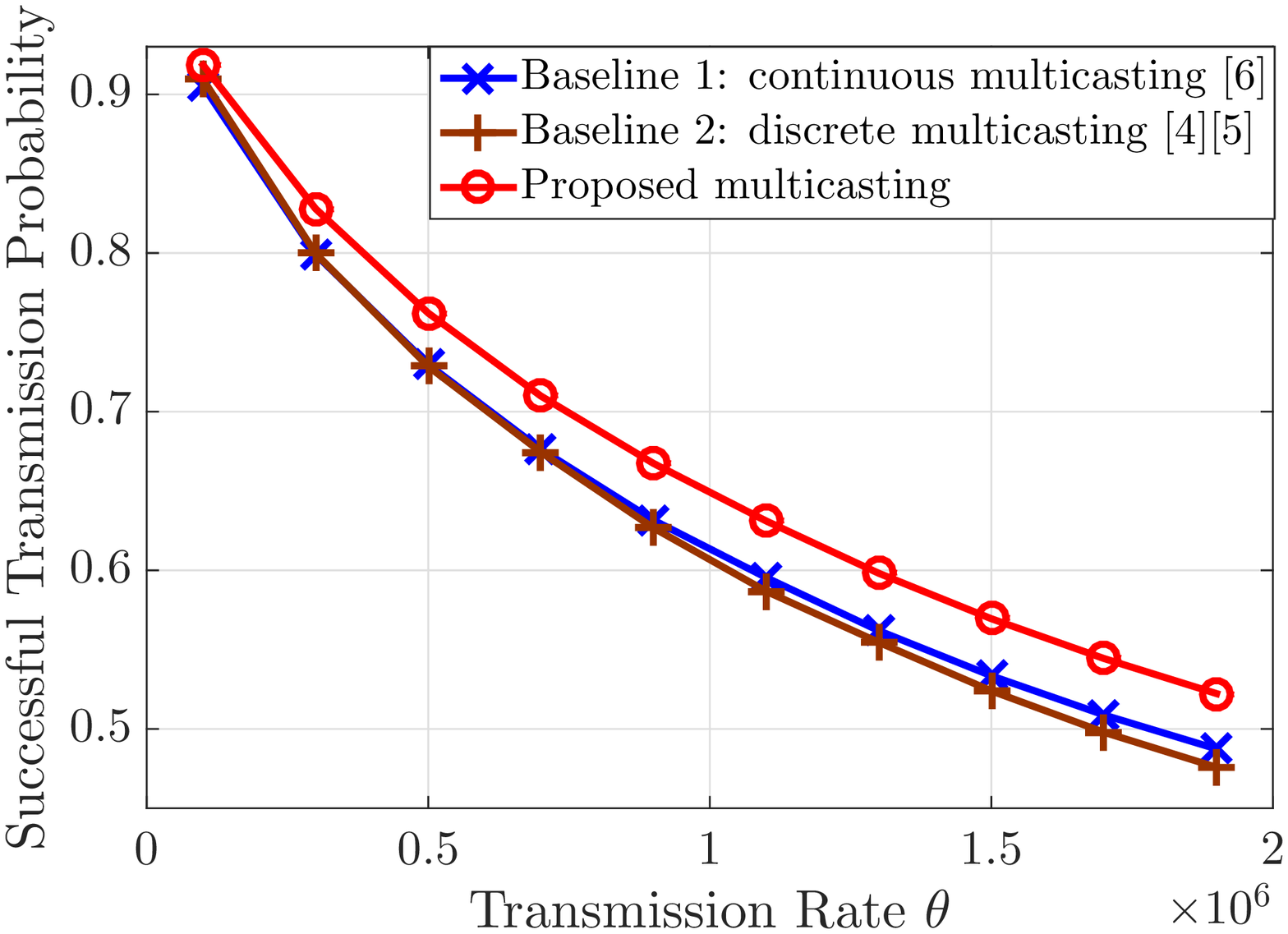}}}\quad\quad
  \subfigure[\label{fig:pro_T_nonmonote}$q({\mathbf p},T)$ versus $T$.]
      {\resizebox{5cm}{!}{\includegraphics{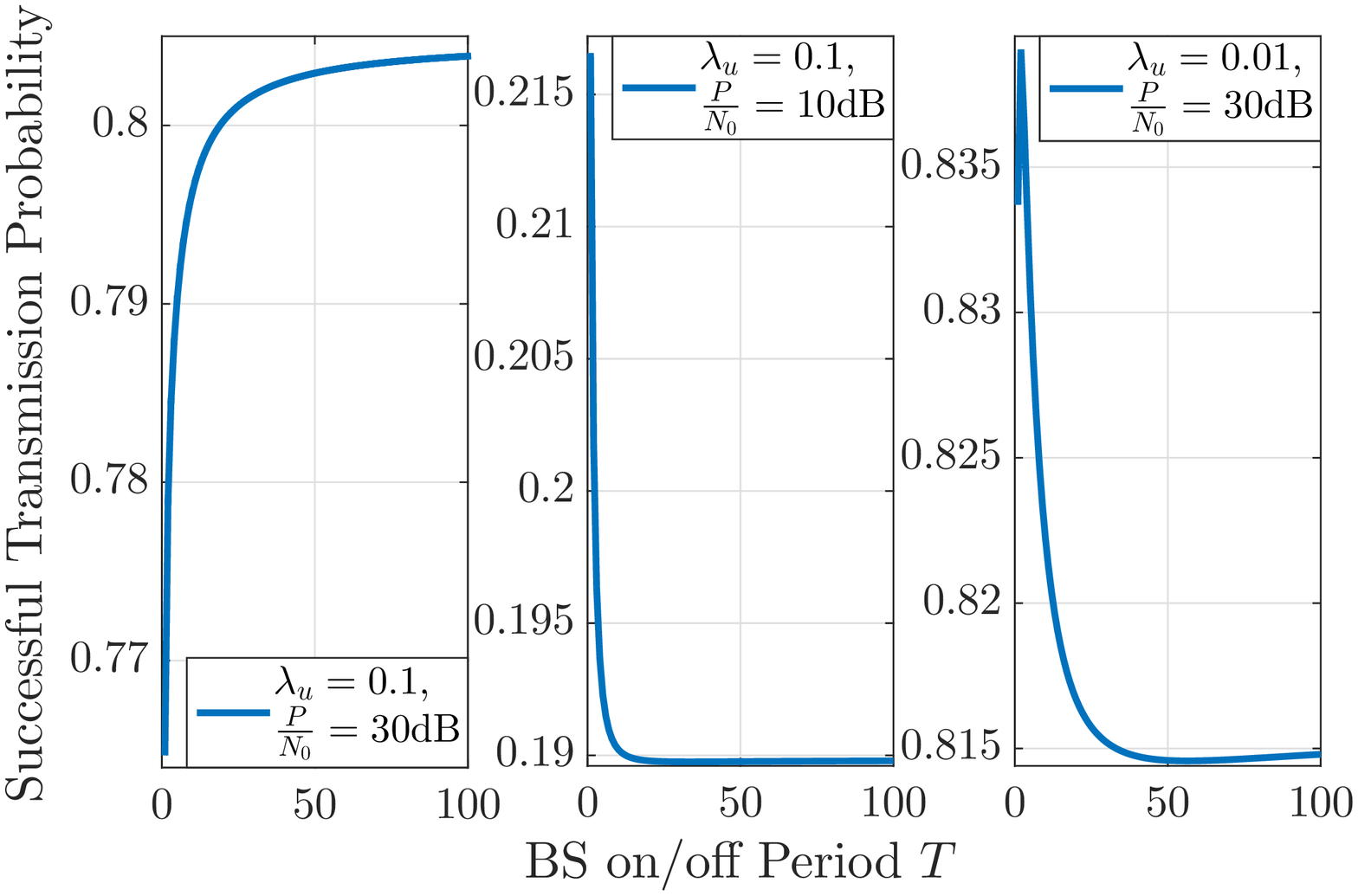}}}
\end{center}
         \caption{\
         Successful transmission probability at  $\lambda_b=0.01$,
         $N=5$, $K=4$,
         $W=10\times10^6$, $\alpha=4$,
         ${\mathbf p}=(0.7, 0.2, 0.06, 0.02, 0.02)$
         and $a_n=\frac{n^{-\gamma}}{\sum_{n\in{\mathcal N}}n^{-\gamma}}$ with $\gamma=2$.
         }\label{fig:simu} \hrulefill
\end{figure*}

In this part, we would like to analyze the successful transmission probability in the general (BS on/off period and user density) region, using tools from stochastic geometry.
For the tractability of the analysis, as in \cite{cui2015analysis}, the dependence between $K_{n,0}$ and ${\rm SINR}_{n,0}$ is ignored.
Then, \eqref{eqn:succ-prob-n-def} can be rewritten as $q_{n}\left({\mathbf p},T\right)=\sum_{k=1}^K\Pr \left[K_{n,0}= k\right]{\Pr}\left[{\rm SINR}_{n,0}\geq2^{\frac{k\theta}{W}}-1\right].$

To obtain $q_{n}\left({\mathbf p},T\right)$,
we first calculate the p.m.f. of $K_{n,0}$, which is denoted as $g_n\left(k,{\mathbf p},T,\lambda_u\right)$.
In calculating this p.m.f., different from \cite{cui2015analysis}, we need the density of the users who request file $m\in {\mathcal N}_{i,-n}$ within the latest $T$ slots up to slot $t_{0}^{'}$, i.e., $\left(1-\left(1-a_{m}\right)^{T}\right)\lambda_u$, instead of the density of the users who request file $m\in {\mathcal N}_{i,-n}$ within one slot, i.e., $a_{m}\lambda_u$.
Next, we calculate the cumulative distribution function (c.d.f.) of ${\rm SINR}_{n,0}$, i.e., ${\Pr}\left[{\rm SINR}_{n,0}\geq\eta\right]$, which is denoted as $f\left(\eta,T_{n},T\right)$.
Note that different from \cite{cui2015analysis}, to evaluate the interference under the temporal-spatial aggregation-based multicasting scheme with periodic BS on/off, we only consider the active BSs in $\Phi_{b,\tau_0}$ at slot $t_{0}^{'}$.
Therefore, we can derive  $q(\mathbf p,T)$.

\begin{Thm}\label{Thm:generalKmulti}
We have $q(\mathbf p,T)
=\sum_{n\in \mathcal N}a_{n}\sum_{k=1}^Kg_n\left(k,{\mathbf p},T,\lambda_u\right)f\left(\eta,T_{n},T\right)$,  
where $g_n\left(k,{\mathbf p},T,\lambda_u\right)$ is given by \eqref{eqn:K-pmf} (at the top of the next page), $f\left(\eta,T_{n},T\right)$ is given by \eqref{eqn:def-f} (at the top of the next page), $\mathcal{SN}_i^1(k-1)\triangleq \left\{\mathcal N_i^1 \subseteq \mathcal N_{i,-n} :|\mathcal N_i^1|=k-1\right\}$,
$\mathbf T_{i,-n}\triangleq\left(T_m\right)_{m\in \mathcal N_{i,-n}}$,
$W_m(T_m,T)\triangleq1+3.5^{-1}\frac{\lambda_u\left(1-\left(1-a_{m}\right)^{T}\right)}{T_m\lambda_b}$, $G_{n,i}(\mathcal N_i^1,\mathbf T_{i,-n},T)
\triangleq \prod_{m\in \mathcal N_i^1}(1-W_m(T_m,T)^{-4.5})\prod_{m\in {\mathcal N_{i,-n}\setminus \mathcal N_i^1}}W_m(T_m,T)^{-4.5}$, $B^{'}( x,y,z ) \triangleq \int_{z}^{1}u^{x-1}(1-u)^{y-1}{\rm d}u$ and $B(x,y)\triangleq\int_{0}^{1}u^{x-1}(1-u)^{y-1}{\rm d}u$.

\begin{figure*}[!t]
\small{\begin{align} \label{eqn:K-pmf}
g_n\left(k,{\mathbf p},T,\lambda_u\right)=&\sum_{i\in \mathcal I_n}\frac{p_i}{T_n}\sum_{\mathcal N_i^1\in \mathcal{SN}_i^1(k-1) }G_{n,i}(\mathcal N_i^1,\mathbf T_{i,-n},T)
\end{align}} 
\small{\begin{align}\label{eqn:def-f}
f\left(\eta,T_{n},T\right)
&=2\pi\lambda_{b}T_{n}\int_{0}^{\infty}d\exp\left(- \eta d^{\alpha}\frac{N_0}{P}\right)\exp\left(-\frac{2\pi}{\alpha T}T_{n}\lambda_{b}\eta^{\frac{2}{\alpha}}d^{2}B^{'}\left(\frac{2}{\alpha},1-\frac{2}{\alpha},\eta+1\right)\right)\exp\left(-\frac{2\pi}{\alpha T}\left(1-T_{n}\right)\lambda_{b}\eta^{\frac{2}{\alpha}}d^{2}B\left(\frac{2}{\alpha},1-\frac{2}{\alpha}\right)\right)\notag\\
&\quad\quad\quad\quad\quad\quad\times\exp\left(-\pi\lambda_{b}T_{n}d^{2}\right){\rm d}d
\end{align}} \hrulefill
\end{figure*}
\end{Thm}

From Theorem~\ref{Thm:generalKmulti}, we can see that in the general region,
the impacts of network parameters $\alpha$, $W$, $\lambda_u$, $\lambda_b$, $\frac{P}{N_0}$, $\mathbf a$ and the design parameters $\left({\mathbf p}, T\right)$ on $q\left({\mathbf p}, T\right)$ are coupled in a complex manner.
Fig.~\ref{fig:pro_tau} plots the successful transmission probability versus the transmission rate $\theta$. Fig.~\ref{fig:pro_tau} verifies Theorem~\ref{Thm:generalKmulti}, and demonstrates the accuracy of the approximations adopted.

In addition, in Fig.~\ref{fig:baselines}, we compare the successful transmission probability of the proposed temporal-spatial aggregation-based multicasting scheme with two baseline schemes. Baseline~1 is a discrete multicasting scheme similar to those proposed in \cite{huang2016delay} and \cite {poularakis2016exploiting}  which consider   temporal aggregation only.
Baseline~2 is the continuous multicasting scheme proposed in \cite{cui2015analysis}  which considers  spatial  aggregation only and can be treated as a special case of our proposed multicasting scheme by setting $T=1$. From Fig.~\ref{fig:baselines}, we can see that our proposed  scheme achieves better performance than the two baselines, especially when the successful transmission probability is small. The performance gain over the discrete multicasting scheme comes from the  improvement of the spectral efficiency, and the performance gain over the continuous multicasting scheme is due to the fact that more multicasting opportunities are utilized by aggregating asynchronous common requests in time.

In the following, we study the impact of the BS on/off period $T$ on $q\left({\mathbf p,T}\right)$, for any given $\mathbf p$.
We characterize the impact of $T$ on the c.d.f. of $\rm SINR_{n,0}$ and the expectation of $K_{n,0}$.

\begin{Lem}\label{Lem:q-T-monotonic_E_k-k-relationship}
$f\left(\eta,T_{n},T\right)$ is a monotone increasing function of $T$, and ${\mathbb E}\left[K_{n,0}\right]$ is a monotone increasing function of $T$.
\end{Lem}

From Lemma~\ref{Lem:q-T-monotonic_E_k-k-relationship}, we can see that, on average, ${\rm SINR}_{n,0}$  and  $K_{n,0}$ both increase with $T$. Note that the successful transmission probability increases with ${\rm SINR}_{n,0}$, but decreases with $K_{n,0}$.  Thus, it is not obvious how $q\left({\mathbf p},T\right)$ changes with $T$. Fig.~\ref{fig:pro_T_nonmonote} indicates that the successful transmission probability may not always increase with $T$.

\section{Performance Analysis in Asymptotic Regions}\label{Sec:asy_ana}
In this part, to obtain design insights, we analyze the asymptotic successful transmission probability in the large BS on/off period region, the large user density region and the small user density region, respectively, using asymptotic approximation.
In these three regions, the successful transmission probability $q\left({\mathbf p}, T\right)$ increases with $T$. In other words,  in these three promising operating regions,   we can reduce energy and increase throughput simultaneously, at the cost of delay increase.

\subsection{Large BS On/off Period Region}
Define $g_n\left(k,{\mathbf p},\infty,\lambda_u\right)\triangleq
\lim_{T\to\infty}\Pr [K_{n,0} =k]$
and $f\left(\eta,T_{n},\infty\right)\triangleq
\lim_{T\to \infty} {\Pr}\left[{\rm SINR}_{n,0}\geq\eta\right]$.
By taking asymptotic approximations of $g_n\left(k,{\mathbf p},T,\lambda_u\right)$ and $f\left(\eta,T_{n},T\right)$, we have the following result.

\begin{Thm}
As $T\to \infty$, we have $q\left(\mathbf p,T\right)=\sum_{n\in{\mathcal N}}a_n\sum_{k=1}^{K}g_n\left(k,{\mathbf p},\infty,\lambda_u\right)f\left(2^{\frac{k\theta}{W}}-1,T_{n},\infty\right)
-\frac{1}{T}Q_1\left(\mathbf p,\lambda_u\right)+{o}\left(\frac{1}{T}\right)$,
where
$g_n\left(k,{\mathbf p},\infty,\lambda_u\right)
=\sum\limits_{i\in {\mathcal I}_{n}}\frac{p_i}{T_n}\sum\limits_{{\mathcal N}_{i}^{'}\in {\mathcal {SN}}_{i}(k-1)}\prod_{m\in \mathcal N_i^1}\big(1-(1+3.5^{-1}\frac{\lambda_u}{T_m\lambda_b})^{-4.5}\big)$ $
\times\prod_{m\in \mathcal N_{i,-n}\setminus\mathcal N_i^1}(1+3.5^{-1}\frac{\lambda_u}{T_m\lambda_b})^{-4.5}$,
$ f\left(\eta,T_{n},\infty\right)=\frac{{D_{0,0}^{-\alpha}}\left|h_{0,0}\right|^{2}P}{N_{0}}$ and $Q_1\left(\mathbf p,\lambda_u\right)>0$.
\label{Thm:pro-asymp-T-domi}
\end{Thm}

By comparing $g_n\left(k,{\mathbf p},\infty,\lambda_u\right)$ with $g_n\left(k,{\mathbf p},T,\lambda_u\right)$ in (\ref{eqn:K-pmf}), we can see as $T\to\infty$,
$\left(1-\left(1-a_{m}\right)^{T}\right)\lambda_u$ converges to $\lambda_u$, i.e., the probability that any file $m\in {\mathcal N}_{i,-n}$ stored in $B_{n,0}$ is requested at least once within $T$ slots converges to the probability that there exists at least one user in the Voronoi cell of $B_{n,0}$ w.r.t. file $m$.
From Theorem~\ref{Thm:pro-asymp-T-domi}, we can see that as $T\to\infty$, $q\left(\mathbf p,T\right)$ converges to the first term of $q({\mathbf p}, T)$, i.e., $\sum_{n\in{\mathcal N}}a_n\sum_{k=1}^{K}g_n\left(k,{\mathbf p},\infty,\lambda_u\right)f\left(2^{\frac{k\theta}{W}}-1,T_{n},\infty\right)$.
From the second term of $q({\mathbf p}, T)$, we can see that $q({\mathbf p}, T)$ increases with $T$, when $T$ is large enough.

\subsection{Large User Density Region and Small User Density Region}
By taking asymptotic approximations of $g_n\left(k,{\mathbf p},T,\lambda_u\right)$, we have the following result.

\begin{Thm}
As $\lambda_u\to\infty$, we have $q\left(\mathbf p,T\right)=\sum_{n\in \mathcal N}a_n f\left(2^{\frac{K\theta}{W}}-1,T_n,T\right)+Q_2\left({\mathbf p},T\right)\frac{1}{\lambda_{u}^{4.5}}+{o}\left(\frac{1}{\lambda_{u}^{4.5}}\right)$,
\label{Thm:pro-asymp-lambda_u-domi}
where $Q_2\left({\mathbf p},T\right)
>0$.
Furthermore, as $\lambda_u\to 0$, we have $q\left(\mathbf p,T\right)=\sum_{n\in{\mathcal N}}a_n f\left(2^{\frac{\theta}{W}}-1,T_n,T\right)-Q_3\left({\mathbf p},T\right)\lambda_u+{o}\left(\lambda_u\right)$,
where $Q_3(\mathbf p,T)
>0$.\label{Thm:pro-asymp-lambda_u_0-domi}
\end{Thm}

Note that as $\lambda_u\to \infty$,  we can see $f\left(2^{\frac{K\theta}{W}}-1,T_n,T\right)$ only, as  the probability that all the $K$ files at $B_{n,0}$ are requested within $T$ slots converges to $1$.
From Theorem~\ref{Thm:pro-asymp-lambda_u-domi}, we can also see that as $\lambda_u\to\infty$, $q\left(\mathbf p,T\right)$ converges to the first term of $q({\mathbf p}, T)$, i.e., $\sum_{n\in \mathcal N}a_n f\left(2^{\frac{K\theta}{W}}-1,T_n,T\right)$, which increases with $T$ (by Lemma~\ref{Lem:q-T-monotonic_E_k-k-relationship}).
Therefore, when $\lambda_u$ is large enough, $q\left({\mathbf p},T\right)$ increases with $T$.
On the other hand, note that as $\lambda_u\to0$, we can only see $f\big(2^{\frac{\theta}{W}}-1,T_n,T\big)$, as only one file cached at $B_{n,0}$ is requested by $u_0$ at slot $t_{0}^{'}$.
From Theorem~\ref{Thm:pro-asymp-lambda_u_0-domi}, we can also see that when $\lambda_u\to0$, $q\left({\mathbf p},T\right)$ converges to the first term of $q({\mathbf p},T)$, i.e., $\sum_{n\in{\mathcal N}}a_n f\big(2^{\frac{\theta}{W}}-1,T_n,T\big)$ which increases with $T$ (by Lemma~\ref{Lem:q-T-monotonic_E_k-k-relationship}).
Therefore, when $\lambda_u$ is small enough, $q\left({\mathbf p},T\right)$ increases with $T$.

\section{Conclusion}

In this paper, by aggregating asynchronous requests in time and space, we proposed a novel temporal-spatial aggregation-based multicasting scheme achieving a better energy-throughput-delay tradeoff for massive content delivery in a large-scale cache-enabled wireless network.  We analyzed the successful transmission probabilities in the general region and three asymptotic regions, respectively. The analytical results offered a new understanding of  energy-throughput-delay tradeoff. More importantly, the analytical results identified three promising operating regions  where we can  increase throughput and reduce energy simultaneously, at the cost of delay increase.

\end{document}